\documentclass[letterpaper]{sig-alternate-2013}
\usepackage{threeparttable}
\usepackage{booktabs}
\newcommand{\tabincell}[2]{\begin{tabular}{@{}#1@{}}#2\end{tabular}}
\permission{\copyright 2017 International World Wide Web Conference Committee (IW3C2),
published under Creative Commons CC BY 4.0 License.}
\conferenceinfo{WWW 2017 Companion,}{April 3--7, 2017, Perth, Australia.}
\copyrightetc{ACM \the\acmcopyr}
\crdata{978-1-4503-4914-7/17/04. \\
http://dx.doi.org/10.1145/3041021.3054142  \\
\includegraphics[scale=1,bb=0 0 30 30]{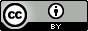}
}
\begin{document}

\setlength{\parskip}{0.5\baselineskip}





%

\title{A Selfie is Worth a Thousand Words: Mining Personal Patterns behind User Selfie-posting Behaviours}

%
%
%
%
%

\numberofauthors{3} 
%
\author{
%
%
\alignauthor
Tianlang Chen\\
       \affaddr{Computer Science}\\
       \affaddr{University of Rochester}\\
       \affaddr{Rochester, NY 14627
}\\
       \email{tchen45@cs.rochester.edu}
\alignauthor
Yuxiao Chen\\
       \affaddr{Computer Science}\\
       \affaddr{University of Rochester}\\
       \affaddr{Rochester, NY 14627
}\\
       \email{ychen211@cs.rochester.edu}
\alignauthor
Jiebo Luo\\
       \affaddr{Computer Science}\\
       \affaddr{University of Rochester}\\
       \affaddr{Rochester, NY 14627
}\\
       \email{jluo@cs.rochester.edu}
}


\maketitle
\begin{abstract}
Selfies have become increasingly fashionable in the social media era. People are willing to share their selfies in various social media platforms such as Facebook, Instagram and Flicker. The popularity of selfie have caught researchers' attention, especially psychologists. In computer vision and machine learning areas, little attention has been paid to this phenomenon as a valuable data source. In this paper, we focus on exploring the deeper personal patterns behind people's different kinds of selfie-posting behaviours. We develop this work based on a dataset of WeChat, one of the most extensively used instant messaging platform in China. In particular, we first propose an unsupervised approach to classify the images posted by users. Based on the classification result, we construct three types of user-level features that reflect user preference, activity and posting habit. Based on these features, for a series of selfie related tasks, we build classifiers that can accurately predict two sets of users with opposite selfie-posting behaviours. We have found that people's interest, activity and posting habit have a great influence on their selfie-posting behaviours. Taking selfie frequency as an example, the classification accuracy between selfie-posting addict and nonaddict can reach 89.36\%. We also prove that using user's image information to predict these behaviours achieve better performance than using text information. More importantly, for each set of users with a specific selfie-posting behaviour, we extract and visualize significant personal patterns about them. In addition, to concisely construct the relation between personal pattern and selfie-posting behaviour, we cluster users and extract their high-level attributes, revealing the correlation between these attributes and users' selfie-posting behaviours. In the end, we demonstrate that users' selfie-posting behaviour, as a good predictor, could predict their different preferences toward these high-level attributes accurately.
\end{abstract}

%
%

%
%

%
%



\section{Introduction}

Popular social media platforms, such as Twitter, Facebook and Instagram, provide their users with a high degree of freedom to express opinions, post interesting images and share immediate news. With the popularity of these platforms among a wide range of users, large amounts of useful information can be mined. Therefore, social media data mining, as a developing research field, has received much attention. A wide spectrum of topics has been studied. For example, in the area of user trait prediction, advanced models are built to predict user information, such as their age \cite{CLU}\cite{JXYH}, gender \cite{DGT}\cite{QSJ}, personality \cite{LDZ}, interest \cite{You2016A} and occupation \cite{THJ}. Also, social media data mining has been applied to a number of meaningful applications, such as disaster management and relief \cite{AGD}\cite{CA}, infectious diseases diffusion analysis \cite{AHV}, election prediction \cite{ATP}\cite{YOU} and street walkability measurement \cite{QAS}.
\begin{figure*}[!htb]
\centering
\includegraphics[scale=0.43]{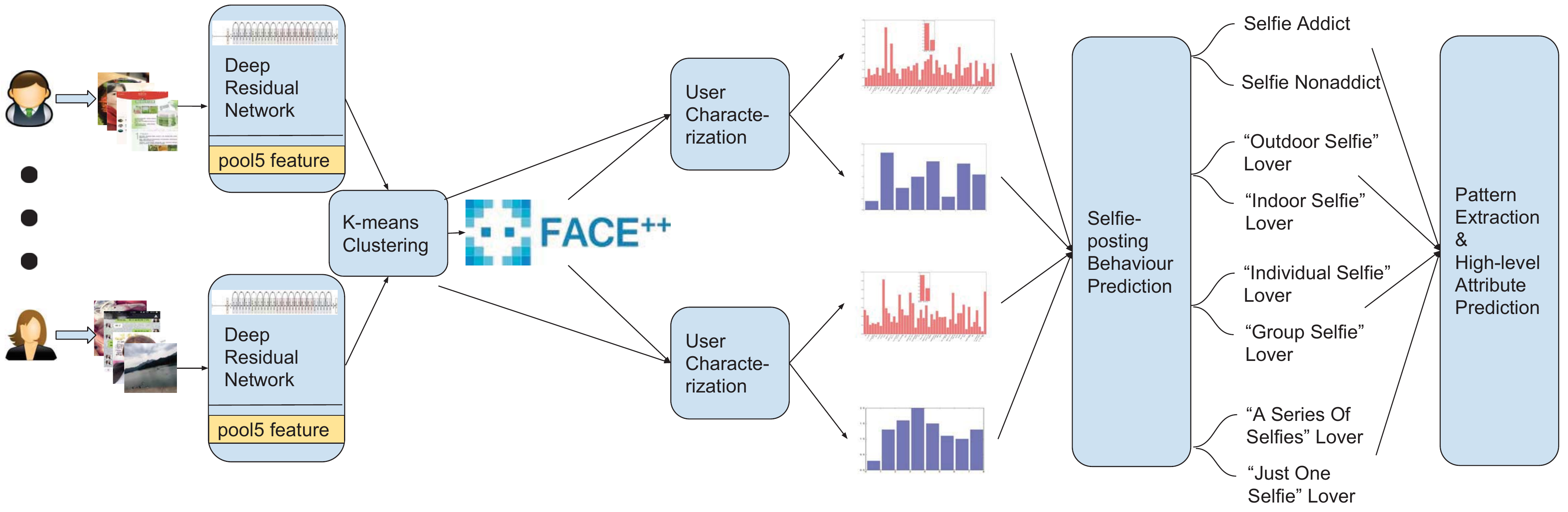}
\caption{Overview of the work.} \label{fig:overview}
\end{figure*}

Selfie-posting behaviour analysis is a less studied topic in computer science. A selfie is a self-portrait photograph, typically taken with a digital camera or camera phone held in the hand or supported by a selfie stick\footnote{https://en.wikipedia.org/wiki/Selfie}. With the popularity of all kinds of social networking services, more and more people choose to post their selfies to their Facebook, Instagram, Twitter and Flicker homepages. They post this type of photos for different reasons, such as recording important events or expressing their moods. Different users have different selfie-posting behaviours, which include comprehensive aspects of selfie-posting information, such as selfie-posting frequency, background selection, with/without other people, and special selfie gestures and preferences. Uncovering the personal patterns behind such diverse information is the motivation of our work. We explore the possibility of predicting whether a person has a specific selfie-posting behaviour from which we attempt to extract user interest, preference, activity and posting habit patterns. The overview of our work is shown in Figure~\ref{fig:overview}.

Our work is based on WeChat Moment. WeChat, developed by Tencent, a leading Internet company in China, is the most extensively used instant messaging platform in China at present. According to the Tencent 2016 Interim Report, the MAU (monthly active users) of WeChat has reached 806 millions. Besides instant messaging, WeChat Moment is one of the most widely used functions provided by WeChat. Like Twitter, WeChat Moment is a platform where users can post text and pictures (up to 9 pictures) in a Moment, which can be accessed and commented by their WeChat friends. However, different from Twitter followers, users' WeChat friends must be approved and therefore are those users who have close relationship with them in real life, such as family members, friends, colleagues and clients. Because of this attribute, comparing with Twitter, Moment enjoys the following advantages: 1) Moments posted by users can reflect users' interest and emotion in a very intimate way as users do not need to worry whether a Moment is appropriate to be seen by unfamiliar people; 2) Moments can directly reflect small social community features.

In this paper, we investigate user selfie behaviors based on their posted images in WeChat Moment. We collect 109,545 images shared by 570 VIP users of a cosmetics brand. The motivation of our work is not only to demonstrate that people's preference and activities can predict their selfie-posting behaviours, but also to establish an explicit relation between people's selfie-posting behaviours and personal patterns. By the end of 2012, Time magazine considered selfie as one of the ``top 10 buzzwords'' of that year, the popularity of selfie creates another to understand people's life style and inner world, this potential makes our work valuable. Our contributions are fourfold:

\begin{list}{$\bullet$}
{ \setlength{\leftmargin}{0.5em}}
    \item We propose a method to accurately classify unlabeled images of WeChat Moment and demonstrate that this approach has a great performance.
    \item We construct significant user-level features and build up classifiers for different selfie-posting behaviour tasks with high accuracy. We uncover diverse interests and activities among users with different selfie-posting behaviours.
    \item We extract users' high-level attributes after clustering them. For each selfie-posting behaviour task, we determine the rank of attributes based on the correlation between the attribute and the corresponding selfie-posting behaviour, which establishes a concise and direct relation between selfie-posting behaviours and user preference.
    \item We demonstrate that users' selfie-posting behaviour have the potential to indicate whether a user has a preference toward a specific high-level attribute.
\end{list}

\section{Related Work}
\textbf{Selfie-posting Behaviour Analysis}. Sharing selfies in social media platforms has become a fashionable behavior among people. It attracts the interests from researchers, especially for psychologists. For instance, Dhir et al. prove that some apparent selfie-posting behavior differences, such as selfie taking frequency and posting frequency, exist between different ages and genders. They demonstrate this by analyzing the questionnaire results from 3763 social media users \cite{DPT}. Qiu et al. notice that some facial expression cues in selfies correlate with human personalities, including agreeableness, conscientiousness, neuroticism, and openness, and can accurately predict openness based on these cues \cite{LJS}. Kim et al. takes advantage of Ajzen's Theory of Planned Behavior to analyze the antecedents of self-posting behavior, finding that users' attitude toward selfie-posting, subjective norm, perceived behavioural control, and narcissism are the significant determinants of an individual's intention to post selfies on SNSs \cite{EJY}.
However, selfie-posting behavior analysis based on the technology of computer vision and machine learning is still missing. To our knowledge, our work is the first research that takes advantage of computer vision and machine learning to analyze and predict human selfie-posting behaviours. Also, compared to the previous research, our current study can obtain more general and objective patterns of users behind their selfie-posting behaviours, because it does not depend on the questionnaires, which are mainly dominated by users' subjective feeling.

\textbf{Deep Residual Networks}. Because of its superior performance in image classification, recognition and segmentation problems, deep convolution neural networks have attracted intense attention. Various advanced network architectures are being proposed constantly \cite{KSH}\cite{SLJ}. Deep Residual Networks \cite{DRN} is one of recently proposed high-performance CNN architectures (winning 2015 ILSVRC \& ROC). It improves performance mainly by letting few stacked layers fit a residual mapping, which exerts the potential of deeper network. Although the network is very deep, it still have lower complexity than VGG nets \cite{SZ}.

\textbf{Mining information from WeChat}. Although most of recent social media data mining researches mainly focus on western social media services, such as Twitter, Facebook as well as Instagram, researchers start to pay attention to WeChat due to its high popularity in China. For example, Qiu et al. analyse the growth, evolution and diffusion patterns of WeChat Messaging Group \cite{JYJ}, Li et al. analyse the diffusions patterns of information in Moments by tracking a large amount of pictures in Moment \cite{ZLY}.

\section{Research Tasks}\label{sec:tasks}
In this paper, we aim to investigate deeper personal patterns behind people's diverse selfie-posting behaviours. To this end, we design several research tasks as follows:

\textbf{R1}: What kinds of people have a preference to post selfies? What kinds of people do not?

\textbf{R2}: What kinds of people have a preference to post a series of selfies in a Moment? What kinds of people have a preference to post only one selfie in a Moment?

\textbf{R3}: What kinds of people have a preference to only post selfies in a Moment? What kinds of people have a preference to post selfies with images of other categories?

\textbf{R4}: What kinds of people have a preference to take selfies with others? What kinds of people prefer to take selfies on his/her own?

\textbf{R5}: What kinds of people have a preference to take selfies outdoors? What kinds of people prefer to take selfies indoors?

\textbf{R6}: What kinds of people have a preference to take selfies with some peculiar behaviours such as holding a gizmo or wearing a facial mask?

\textbf{R7}: Is there a more concise and direct way to indicate the correlation between people's preference and their selfie-posting behaviour? Could we predict people's preference from their selfie-posting behaviour?

For each task, We propose a specific selfie-posting measures to explore it quantitatively. All selfie-posting measures that correspond to these tasks will be defined in Section~\ref{sec:experiments}.

\section{Methodology}
\subsection{Moment Image Classification}
To construct user's profile based on their Moment image, we first classify each image into different categories, so standardized approach can be used to characterize user. Also, the dimension of feature vector will be in a reasonable range. Considering that we do not have any labels for the Moment images, we classify each image by extracting and clustering its deep features. The whole process is as follows. First, we extract deep features from the Deep Residual Network model proposed by He et al. \cite{DRN}. In particular, We extract a deep-level 2048-dimensional feature vector for each image from the last ``pool5'' layer of ResNet-50. After that, we cluster these feature vectors by k-means clustering. We determine the value of k based on Silhouette Coefficient. When computing it, to reduce time complexity, we replace the mean distance of a sample to all samples of a cluster with the distance between this sample and the centroid of this cluster. We set k from 10 to 100 and find that when k is larger than 60, there is a marked decline for the Silhoette Coefficient. So we set K = 60 originally and obtain 60 categories with their corresponding Moment images. Next, we manually combine several categories that we judge to be the same category, generating 47 categories, and label them according to their corresponding images. The names of the 47 categories are shown in Table~\ref{tab:number} and the classification performance evaluation will be recorded in Section \ref{sec:experiments}.

Some well-known social networking services have their official image categories, such as Pinterest\footnote{https://www.pinterest.com/}. For Pinterest, there are a total of 34 available categories for users to choose from. Therefore, we compare our categories with the categories of Pinterest and discuss some different points. First of all, considering that the users of our dataset are all VIP of a cosmetic brand, some of our categories are consistent with these kinds of users' special attributes, such as Cosmetic, Cosmetics Ad and Bracelet \& Necklace. On the other hand, there is a considerable amount of Moment images that are related to WeChat, such as Chat Screenshot, WeChat Moment and WeChat Expression. In addition, other than content, some images the users post have special styles, such as Special Effects Photos and Very Long Pictures (WeChat Moment allows user to post images that is unrestricted in their height), so we also classify them into separated categories. Most importantly, with different scopes of applications for these two services, the categories of Pinterest majorly focus on people's interest, while the categories of WeChat Moment have a wider range of coverage involving users' interest, preference, activity and occupation.

\subsection{User Characterization}\label{sec:features}
After we classify the images in users' Moments, we can characterize each user according to their posted images. We first give some significant definitions based on category. These definitions will be used to characterize users.

\newdef{definition}{Definition}
\definition{(\textbf{Occurrence of a category})}. An occurrence of a category is defined as the existence of image in a Moment that belongs to this category, regardless of the total image number of this category in this Moment, it is regarded as an occurrence of this category.

\definition{(\textbf{Frequency of a category for a user})}. Generally, a user has a number of Moments. For a user, his/her posting frequency of a category is defined as the total occurrence number of this category for these Moments divided by the total occurrence number of all categories for these Moments. It reflects whether a user has a preference to post a specific category in a Moment.

\definition{(\textbf{Inertia of a category for a user})}. For a user, his/her posting inertia of a category is defined as the total image number of this category in all his/her Moments divided by the total occurrence number of this category in all his/her Moments. It reflects whether a user has a preference to posting relatively more images of a specific category in a Moment.

\definition{(\textbf{Singleness of a category for a user})}. For a user, his/her posting singleness of a category is defined as the number of Moments that only have occurrence of this category divided by the number of Moments that have occurrence of this category. It reflects whether a user has a preference to only including a single specific category in a Moment.

We intend to exploit the relationship between users' image-posting behaviour of other categories and their selfie-posting behaviours, as we think people's interests, activities and posting habits have a large effect on their selfie-posting behaviours. We make sure that user features do not include any user selfie information. As a result, we extract three kinds of features as follows:

\begin{list}{$\bullet$}
{ \setlength{\leftmargin}{0.5em}}
    \item User's frequency feature (F-feature). For a user, it is a combination of the frequencies of all categories except selfie. We compute the frequency of each category after filtering out all selfie images of this user. As there are a total of 46 categories besides selfie, this feature is a 46-dimensional vector, which is related to user's interests and activities.
    \item User's inertia feature (I-feature). For a user, we compute it as the total image number of all the other categories in all his/her Moments divided by the total occurrence number of all the other categories in all his/her Moments. This feature is a value related to user's posting habit.
    \item User's singleness feature (S-feature). For a user, we compute it as the total number of Moments that only contain one category divided by the total number of Moments after filtering out all Moments that includes selfie images. This feature is a value related to user's posting habit.
\end{list}

Eventually, each user can be characterized as a 48-dimensional feature vector (46+2), we will illustrate how to use these feature vectors to perform the selfie-related research tasks mentioned in Section~\ref{sec:tasks}.

\section{Experiments}\label{sec:experiments}

\subsection{Dataset}

We collect a dataset from WeChat Moment, which consists of 570 users with their 37,359 Moments and 109,545 Moment images, from Mar 21, 2016 to July 21, 2016. All of these users are VIP of a cosmetic brand. This kind of users' Moments include a considerable amount of selfie images, as well as a sufficient number of images for each subcategory of selfies. We check the dataset and find that almost all the users are female and between about 20 and 40 years old. This fairly targeted datset helps reduce the influence of irrelevant variables so that we will extract more credible information about the relation between people's selfies and their life details.

\subsection{Experiment Results}

\begin{figure}[!htb]
\centering
\includegraphics[scale=0.09]{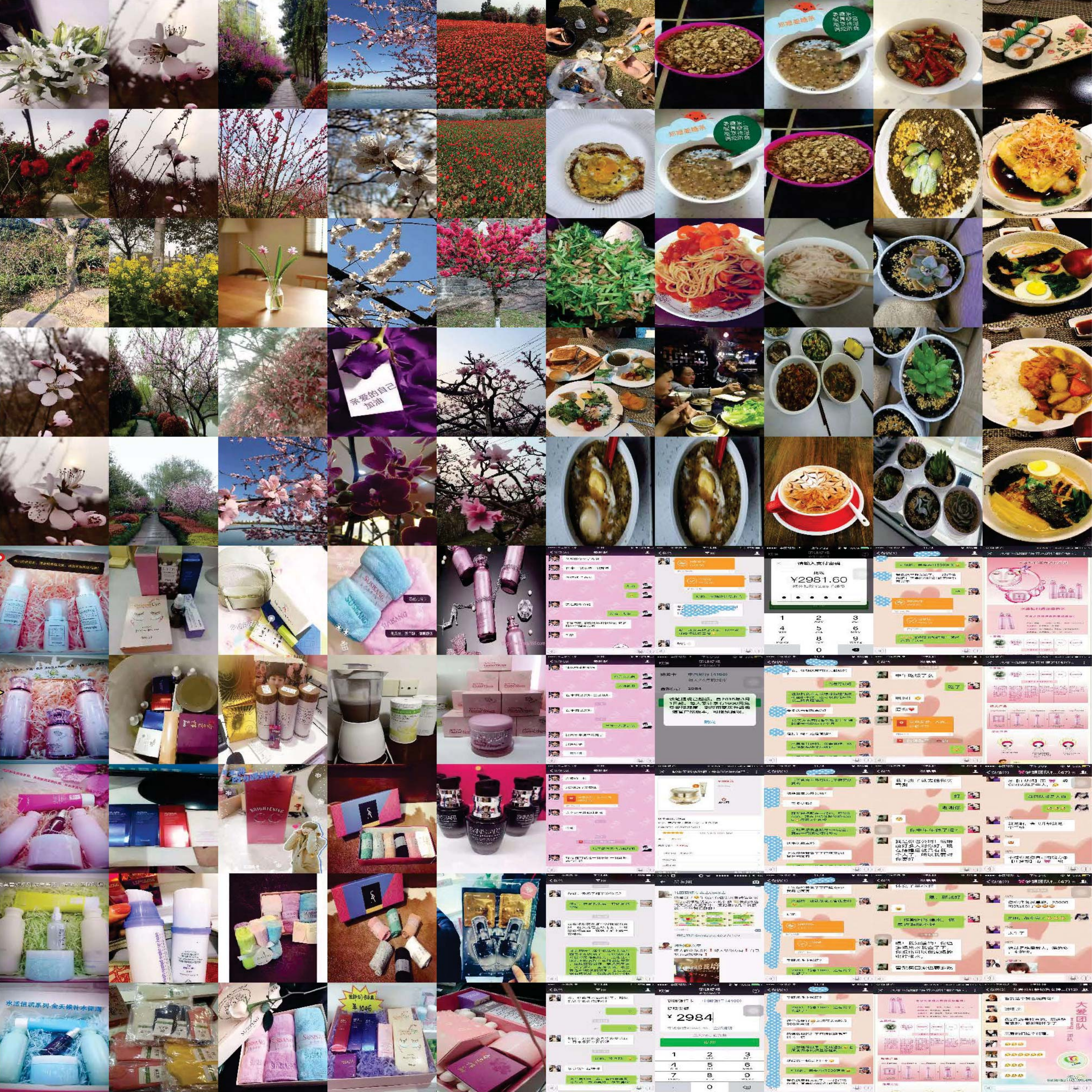}
\caption{The clustering results for several typical categories. From upper left to bottom right: Flower, Meal, Cosmetic, and Chat Screenshot.} \label{fig:imageclass1}\vspace{-1mm}
\end{figure}

Table~\ref{tab:number} shows the total image number in each category after we classify images. We can see that Selfie as the most dominant category, its proportion is more than 10\%. To evaluate the performance of the image classification approach, for each category, we randomly sample 500 images and let 2 volunteers to judge whether it is accurate to classify an image into this category. The average accuracy for all categoires is 88.5\%, standard deviation is 9.12 and 39 of 47 categories are higher than 80\%, We show the classification results for several typical categories in Figure~\ref{fig:imageclass1}. We can see that our unsupervised image classification method produces very coherent categories.
\begin{table}[htbp]
  \small
  \caption{Total image number of each category}
  \begin{threeparttable}
   \scalebox{0.9}{
    \begin{tabular}{|c|c|c|c|}

    \cline{1-4}
    Name & Num. &Name & Num. \\ \cline{1-4}
    Indoor Selfie & 3222 &Cosmetic Tips&3952\\ \cline{1-4}
    Pet & 1180&Display Rack&1301\\ \cline{1-4}
    Bed &2083&Hand \& Leg&2197\\ \cline{1-4}
    Big Word Ad &1239&Wallet \& Accessory&1949\\ \cline{1-4}
    Small Group Photo &968&Fruit \& Cake&1973\\ \cline{1-4}
    Poster &1922&WeChat Moment&1989\\ \cline{1-4}
    Chart &1157&Motto&984\\ \cline{1-4}
    Pink Goods &1431&WeChat Expression&1281\\ \cline{1-4}
    Child &5491&QR-code&1281\\ \cline{1-4}
    Flower &2105&WeChat Wallet&1891\\ \cline{1-4}
    Cosmetic &3878&Chat Screenshot&5506\\ \cline{1-4}
    Cosmetics Ad &1711& Other Ad&1029\\ \cline{1-4}
    Activity &1544&Comic&1767\\ \cline{1-4}
    Large Group Photo &1195&Essay&2081\\ \cline{1-4}
    Building &2548&Other Goods&2612\\ \cline{1-4}
    TV \& Poster Screenshot &1818&Shoes&2538\\ \cline{1-4}
    Toy &2022&Necklace \& Bracelet&2491\\ \cline{1-4}
    Snack&1731&Clothes&1855\\ \cline{1-4}
    Landscape Photo&2519&Baby&1367\\ \cline{1-4}
    Tourist Photo&3567&Full-Length Photo&5183\\ \cline{1-4}
    Sunglass \& Handbag&3279&Special Effects Photo&2424\\ \cline{1-4}
    Photoshop Photo&1439&Very Long Picture&245\\ \cline{1-4}
    Star&1959&Meal&2977\\ \cline{1-4}
    Beauty Ad &1951&Outdoor Selfie &2359\\ \cline{1-4}
    Holding Something Selfie &3032&Face Mask Selfie &1467\\ \cline{1-4}
    Total &109545 & & \\ \cline{1-4}

    \end{tabular}}%
    \tiny * ``Tourst Photo'' represents full-length photo of tourist in a travel background. ``Photoshop Photo'' represents photo with words/graphs using photoshop.
    \end{threeparttable}
  \label{tab:number}%
\end{table}%

\begin{figure}[!htb]
\centering
\includegraphics[scale=0.25]{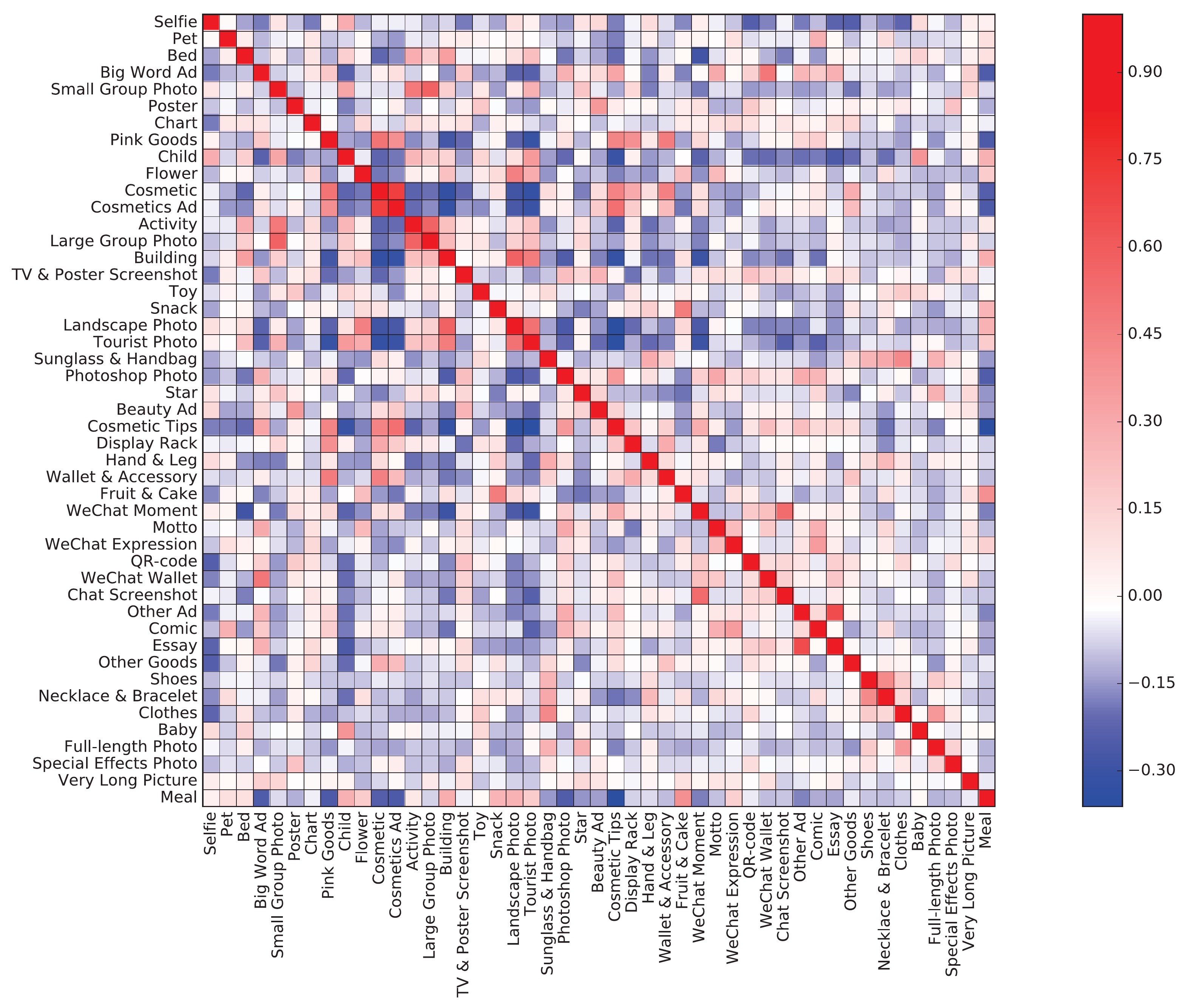}
\caption{Pearson correlation coefficients between different categories.} \label{fig:pearson}\vspace{-1mm}
\end{figure}

We calculate each user's frequency distributions of all categories and compute the Pearson correlation coefficient between each category as shown in Figure~\ref{fig:pearson}. The result is consistent with our expectation. High Pearson correlation coefficients exist between some related or similar categories, such as Cosmetic and Cosmetics Ad, Building and Tourist Photo, Meal and Fruit \& Cake, Clothes and Sunglass \& Bag, and so on. On the other hand, negative Pearson correlation coefficients exist between some categories, such as Building and Cosmetic, Landscape Photo and Cosmetics Ad, Tourist Photo and WeChat Moment, and so on. For the Selfie category, it has relatively high correlation coefficients with Child, Baby, Star, Beauty Ad, Tourist Photo, and Meal.

Before we perform the research tasks outlined in Section~\ref{sec:tasks}, we first filter out the users whose total occurrence number of all categories is lower than 50. This pre-processing avoids the inaccuracy in a user's frequency distributions due to sparse data. After this process, our dataset includes 283 users, 33,877 Moments and 100,677 images.

\subsubsection{Tasks R1 - R3}\label{sec:R1-R3}

To perform R1 - R3, we first design a prediction task to investigate whether people's interest, activity and posting habit can determine these basic selfie features. For research task R1, we sort the users by their selfie frequency, labeling the top 25\% users as positive and the bottom 25\% users as negative. This task can be regarded as a binary classification task. For research task R2 and reseach task R3, we first filter out the bottom 1/3 users sorted by their total occurrence number of selfies, since selfie inertia and selfie singleness will be invalid if the total selfie occurrence number is too low. After that, we sort the users by their selfie inertia and selfie singleness, respectively, and label them in the same way as task 1. We select different features and fusion strategies for each user according to Section~\ref{sec:features}, and for each task, we perform a 10-fold cross-validation using SVM. In the end, the 10-fold cross-validation result is shown as Table~\ref{tab:r1r2r3}.

\begin{table*}[htbp]\centering
  \small
  \caption{\label{tab:r1r2r3}10-fold cross-validation result of selfie frequency, selfie inertia and selfie singleness.}
  \begin{threeparttable}

  \begin{tabular}{|r|r|r|r|r|r|r|r|}

    \cline{1-8}
    Classfication Accuracy & F-feature & I-feature &S-feature&F+I-feature&F+S-feature&S+I-feature&F+S+I-feature\\ \cline{1-8}
    Selfie Frequency&\textbf{89.36}&/&/&/&/&/&/\\ \cline{1-8}
    Selfie Inertia&65.84 &73.63&64.08&\textbf{83.52}&/&74.76&77.67\\ \cline{1-8}
    Selfie Singleness &61.54&65.93&\textbf{77.67}&/&\textbf{77.67}& 70.87&72.82\\ \cline{1-8}

    \end{tabular}%

    \small * "/" indicates either it makes no sense or will seriously decrease the accuracy.

  \end{threeparttable}
\end{table*}%

From Table~\ref{tab:r1r2r3}, we can see that for classification of two sets of users (users labeled as positive and users label as negative) with different selfie frequency, the 10-fold cross-validation accuracy reaches 89.36\%, which proves that a user's posting frequency of other categories can accurately predict whether this user is a selfie addict or not. For classification of two sets of users with different selfie inertia, the 10-fold cross-validation accuracy reaches  83.52\% using the fusion of user's frequency feature and inertia feature. It shows that a user's selfie inertia is related to his/her posting frequency of other categories, but the most important feature is the user's inertia feature. A similar conclusion can be drawn for classification of two sets of users with different selfie singleness, the accuracy is 77.67\% using the fusion of user's frequency feature and singleness feature.

To demonstrate that users' image information is more effective to predict their selfie-posting behaviour, we also extract their text information as a comparison. We still implement the binary classification task for user's selfie frequency in the same way as task R1, but based on user's text information. Each user have corresponding Moments and a Moment includes a text. So we design two approaches for the task. For the first one, we train an Long Short-Term Memory (LSTM) network with word embedding to predict whether a text belongs to a Moment that contains Selfie. We label a text as positive if its corresponding Moment contains image(s) that belongs to Selfie category. For each text, the network output has 2 nodes which record the probability of positive and negative. Then we compute each user's mean positive probability of all his/her texts. In the end a threshold will be learned from users in the training set to determine whether a value of mean probability will be classified as positive or negative. All texts that belong to users whose selfie frequency rank is between 25\% to 75\% are used to train/evaluate the LSTM network. For other users, we randomly select 70\% of them as training samples to learn the threshold and 30\% of them as test samples. We label users in the same way as task R1. The accuracy of LSTM to classify a text is only 59.32\%, and the classification accuracy on user level is 67.86\%. The second approach is to implement doc2vec on each text, and compute the mean vector for all texts of a user to represent his/her text feature. For this, the accuracy is below 55\%. This two approaches demonstrate that using user's image information is more effective.

\begin{figure}[!htb]
\centering
\includegraphics[scale=0.17]{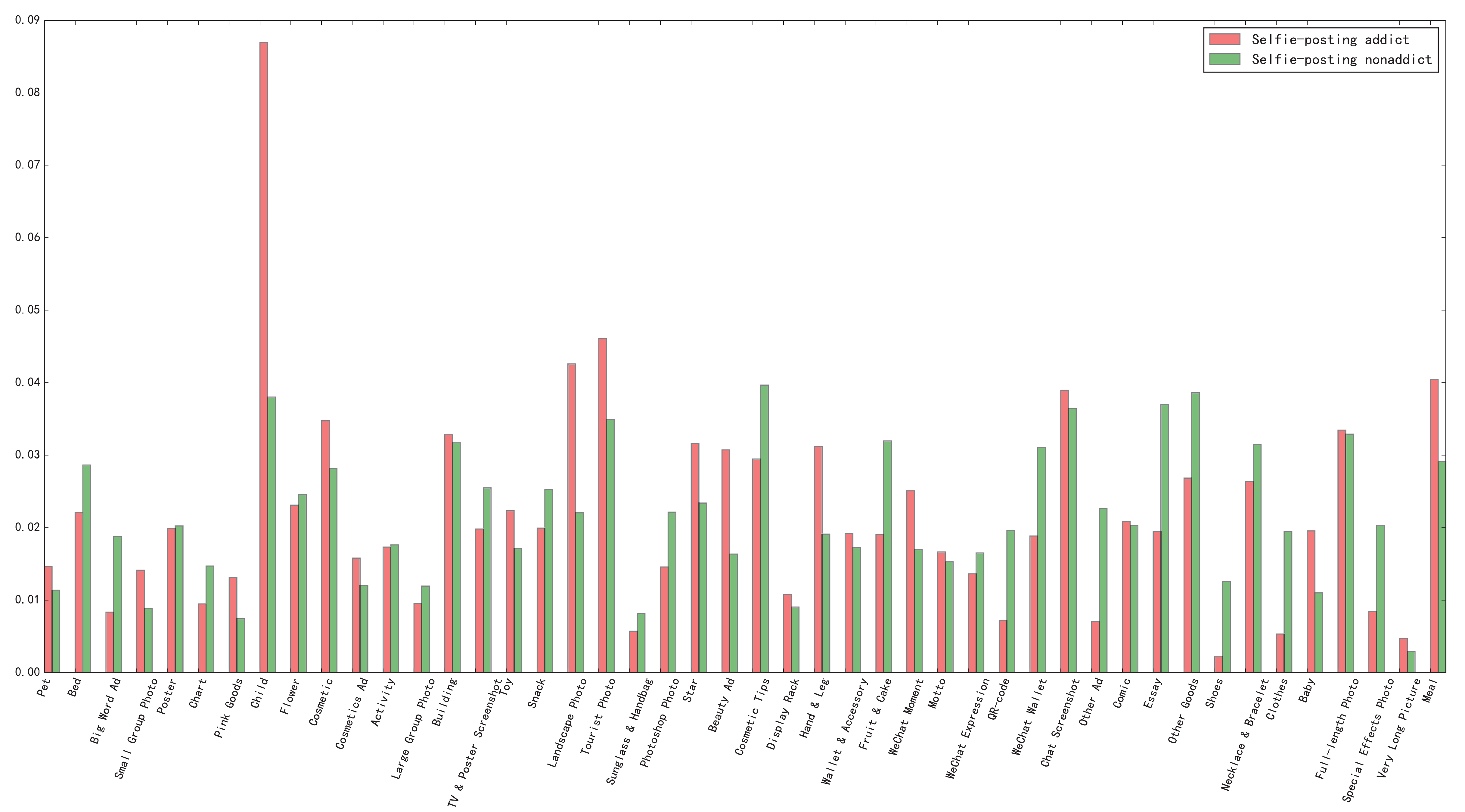}
\caption{Comparison between selfie-posting addict and selfie-posting nonaddict.} \label{fig:r1}\vspace{-1mm}
\end{figure}

\begin{figure}[!htb]
\centering
\includegraphics[scale=0.17]{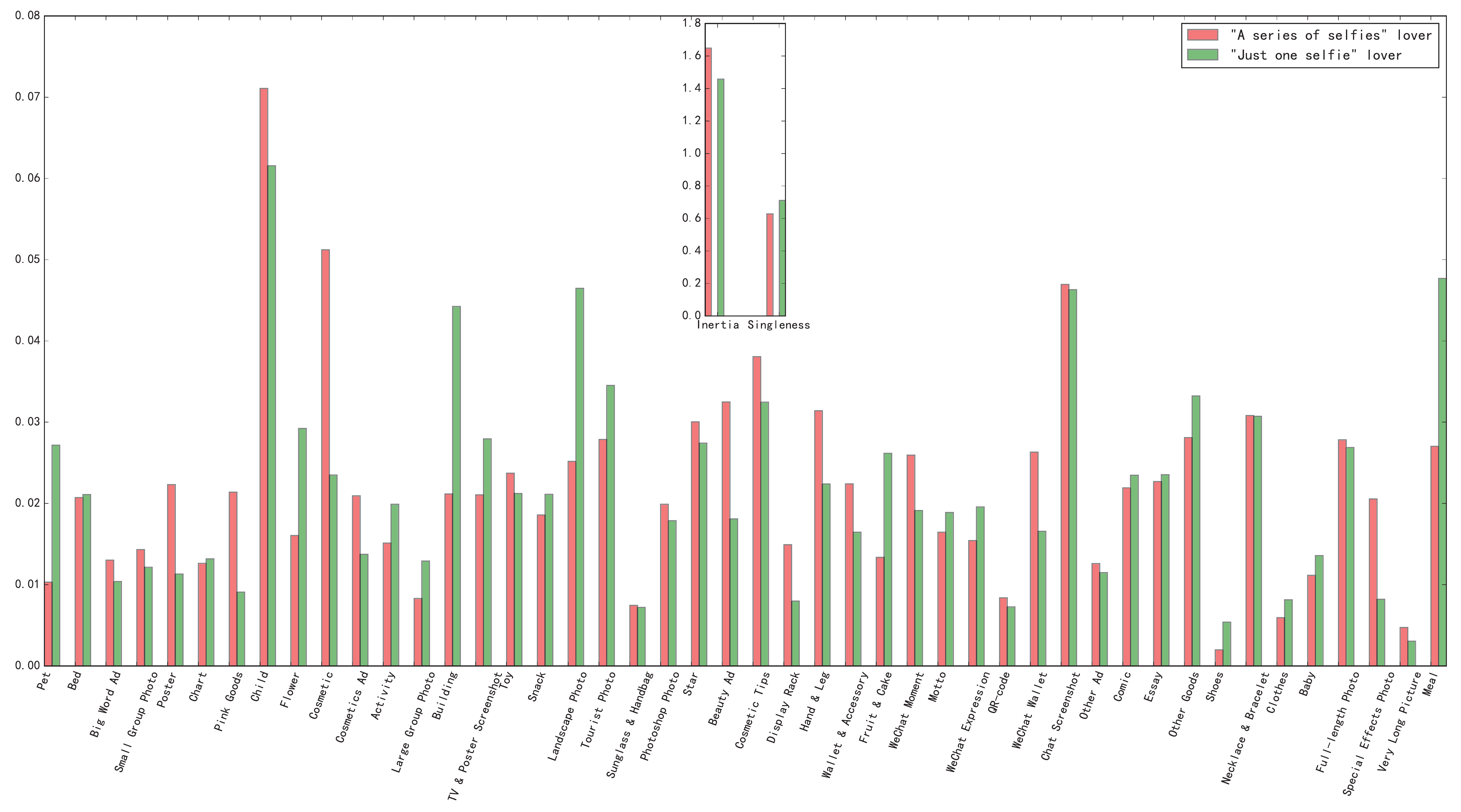}
\caption{Comparison between ``a series of selfie'' lover and ``just one selfie'' lover.} \label{fig:r2}\vspace{-1mm}
\end{figure}

\begin{figure}[!htb]
\centering
\includegraphics[scale=0.17]{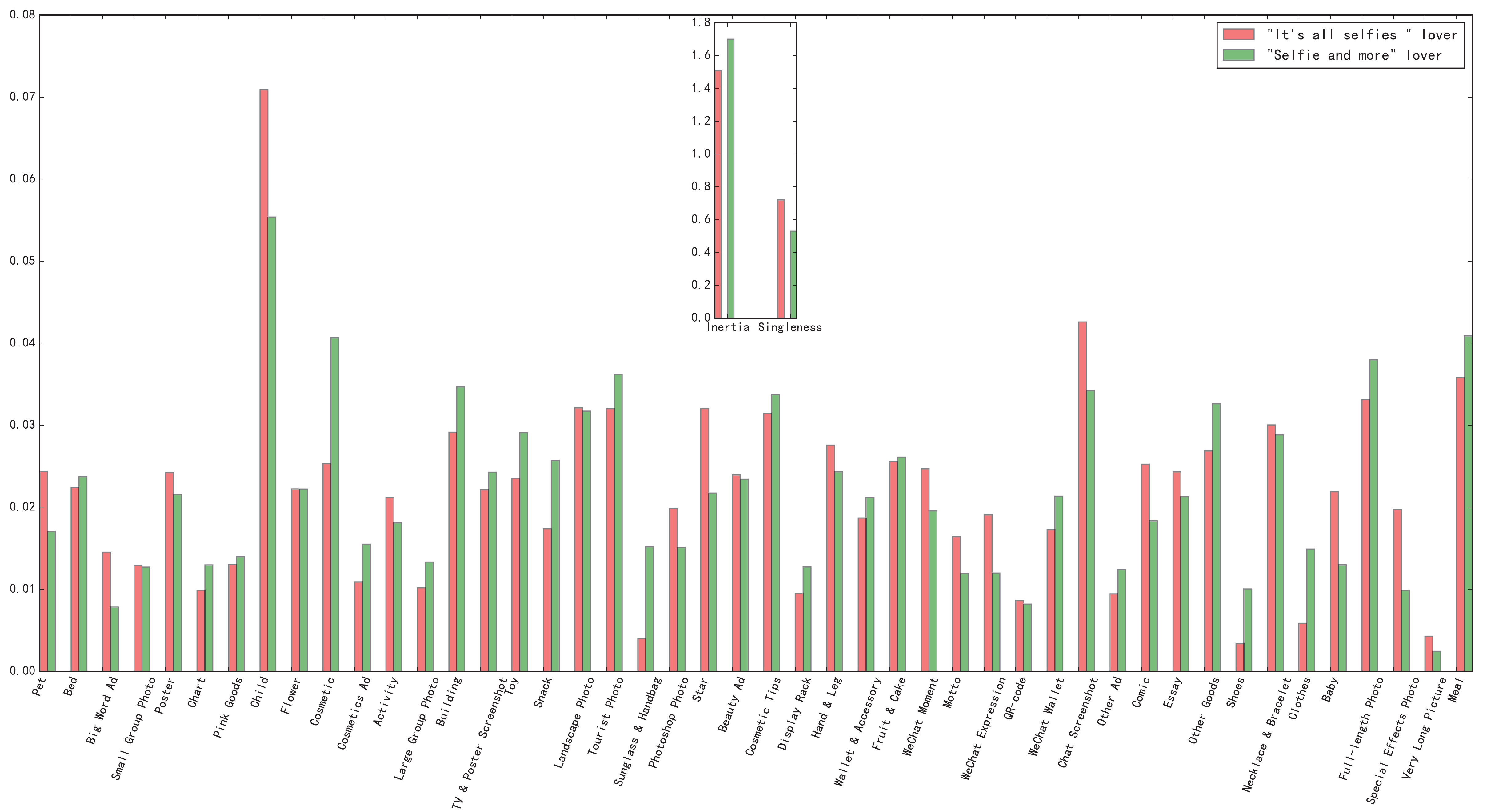}
\caption{Comparison between ``It's all selfies' lover and ``selfie and more'' lover.} \label{fig:r3}\vspace{-1mm}
\end{figure}

For each task, to further investigate the difference between two sets of users and reveal the personal patterns behind a specific selfie-posting behaviour, we respectively compute the mean value of each feature on all users in each user set. Figure~\ref{fig:r1} - Figure~\ref{fig:r3} show the comparison results between two sets of users for each task.

Users' posting preference reflects their interests and activities. From Figure~\ref{fig:r1}, we can see that selfie-posting addicts have a clear preference to post images about Child (a picture taken with children is also included in this category). Also, they are more likely to post images about Cosmetic, Cosmetics Ad, Pink Goods, Landscape Photo, Tourist Photo, Hand \& Leg, Star, Beauty Ad, and Meal. In contrast, nonaddicts are more interested in posting images about Shoes, Clothes, Chart, Other Ad, Fruit \& Cake, and Essay.

As shown in Figure~\ref{fig:r2}, whether a user have a preference to post a large number of selfies in a Moment is related to his/her posting frequency of other categories. ``A series of selfies'' lovers are more likely to post images about Pink Goods, Cosmetic, Hand \& Leg, Display Rack, Special Effects Photo, and Child, while the opposite users are more likely to post images about Building, Tourist Photo, Meal, Pet, Flower and Fruit \& Cake. However, the main factor is a user's posting habit. As the sub-figure of Figure~\ref{fig:r2} shows, ``A series of selfies'' lovers tend to have higher inertia feature, which reveals that other than selfies, they are also accustomed to posting a great number of images that belong to the same category in a Moment. In addition, their singleness feature is lower, which means that they love to include more than one categories in a Moment.

From Figure~\ref{fig:r3}, it can be seen that whether a user have a preference to only post selfies in a Moment is related to this user's  posting frequency of other categories. ``It's all selfies'' lovers are more likely to post images about Child, Star, Chat Screenshot, and Wechat Expression. The opposite users have a higher possibility to post images about Building, Snack, Clothes and Sunglass \& Beg. Still, the posting habit is the paramount factor. ``It's all selfies'' lovers usually have a higher value for the singleness feature and lower value for the inertia feature.


\subsubsection{Tasks R4 - R6} \label{sec:R5-R7}

\begin{figure}[!htb]
\centering
\includegraphics[scale=0.6]{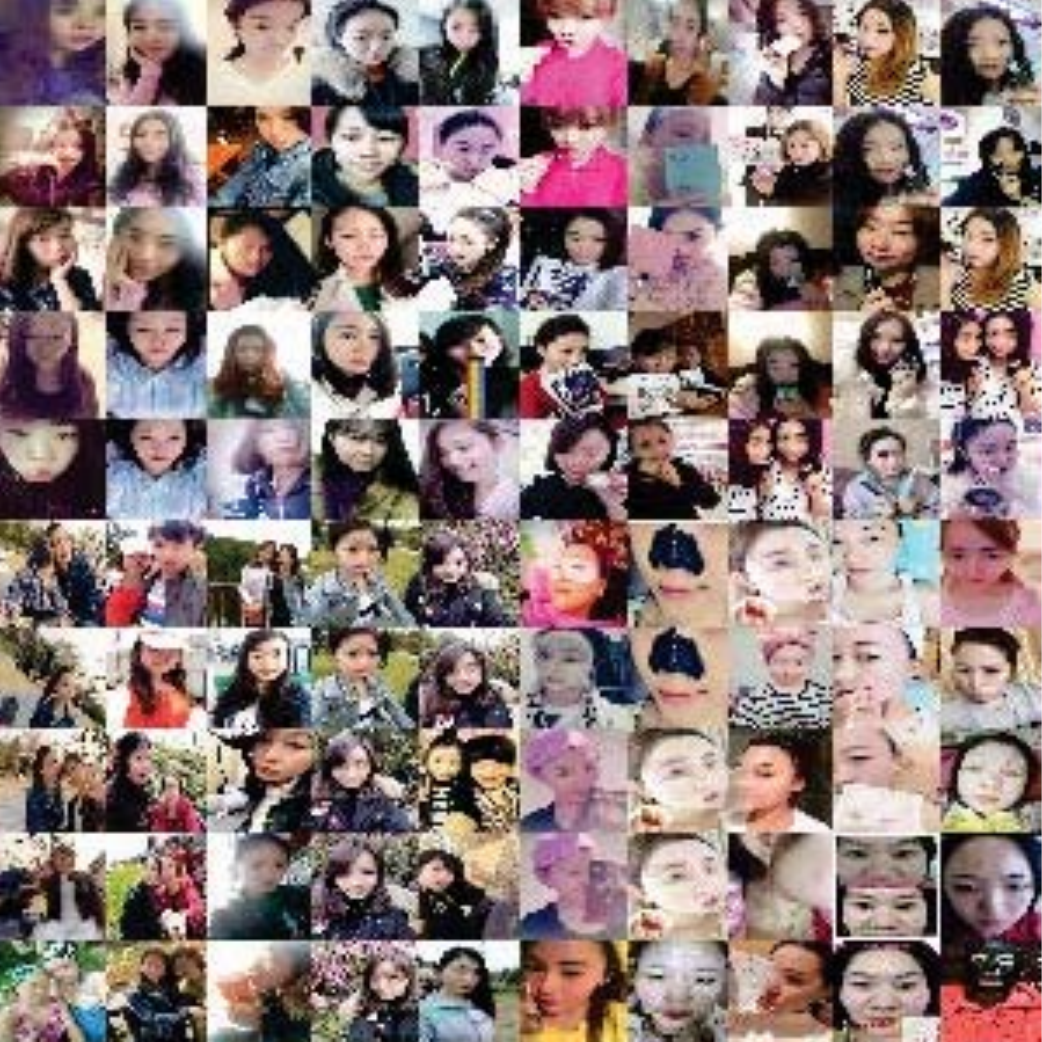}
\caption{The clustering results for Selfie subcategories. From upper left to bottom right: Indoor Ordinary Selfie,  Holding Something Selfie, Outdoor Selfie, Face Mask Selfie.} \label{fig:imageclass2}\vspace{-1mm}
\end{figure}

To investigate advanced selfie-posting tasks mentions in task R4 - task R6, we classify selfie images in two different directions. On one hand, we continuously implement K-means clustering to cluster the deep-level 2048-dimensional features into 4 subcategories, including Indoor Ordinary Selfie, Outdoor Selfie, Holding Somthing Selfie and Face Mask Selfie, the total image number of each subcategory is shown in Table~\ref{tab:subcategory1}. We show the clustering result of each subcategory in Figure~\ref{fig:imageclass2}. On the other hand, we detect the number of faces in each image using Face++\footnote{http://www.faceplusplus.com} and classify each image into One-face Selfies / Multi-face Selfies, the total image number of each subcategory is also shown in Table~\ref{tab:subcategory1}.
\begin{table}[htbp]
  \centering\small
  \caption{The names and numbers of subcategories of selfies.}
    \begin{tabular}{|r|r|r|r|}

    \hline
    \multicolumn{2}{|c|}{Feature Extraction + K-means}&\multicolumn{2}{|c|}{Face++}\\ \hline
    Indoor Ordinary Selfie &3222 &Individual Selfie &8960 \\ \hline
    Outdoor Selfie & 2359 &With Other Selfie & 1120\\ \hline
    Holding Something Selfie & 3032 & & \\ \hline
    Face Mask Selfie & 1467 & & \\ \hline
    Total & 10080 & Total & 10080 \\ \hline

    \end{tabular}%
  \label{tab:subcategory1}%
\end{table}%

We define four advanced selfie-posting measures for task R4 - task R6 as follows:

\definition{(\textbf{Group selfie tendency for a user})}. \label{def:1}
For a user, his/her group selfie tendency is defined as the total occurrence number of the Multi-face Selfie subcategory divided by the sum of occurrence number of One-face Selfie subcategory and Multi-face Selfie subcategory.

\definition{(\textbf{Outdoor selfie tendency for a user})}. \label{def:3}
For a user, his/her outdoor selfie tendency is defined as the total occurrence number of the Outdoor Selfie subcategory divided by the sum of occurrence number of Outdoor Selfie subcategory and Indoor Ordinary Selfie subcategory.

\definition{(\textbf{Holding something selfie tendency for a user})}. \label{def:4}
For a user, his/her holding something selfie tendency is defined as the total occurrence number of the Holding Something Selfie subcategory divided by the sum of occurrence number of Holding Something Selfie subcategory and Indoor Ordinary Selfie subcategory.

\definition{(\textbf{Face mask selfie tendency for a user})}.\label{def:2}
For a user, his/her face mask selfie tendency is defined as the total occurrence number of Face Mask Selfie subcategory divided by the sum of occurrence number of Face Mask Selfie subcategory and Indoor Ordinary Selfie subcategory.

Definitions~\ref{def:1}, ~\ref{def:3} respectively correspond to task R4, R5 and Definitions~\ref{def:4}, ~\ref{def:2} correspond to task R6. For each task, to avoid the inaccuracy due to sparse data, we first filter out the bottom 25\% user who have lowest sum of occurrence number of the corresponding two subcategories. Then we sort the users by the tendency of each task, labeling the top 25\% users as positive and bottom 25\% users as negative. We only select the user frequency feature since other features are useless and may adversely affect the results. The 10-fold cross-validation accuracy for each task is shown in Table~\ref{tab:r5r6r7}.

\begin{table}[htbp]
  \centering\small
  \caption{\label{tab:r5r6r7}10-fold cross-validation results of selfie tendencies.}
  \begin{threeparttable}
    \begin{tabular}{|r|r|r|r|}

    \cline{1-2}
    Advanced Selfie Tendency & Classification Result \\ \hline
    Group Selfie  & 78.64 \\ \hline
    Outdoor Selfie & 80.20 \\ \hline
    Holding Something Selfie & 92.55 \\ \hline
    Face Mask Selfie & 90.32 \\ \hline

    \end{tabular}%

  \end{threeparttable}
\end{table}%

\begin{figure}[!htb]
\centering
\includegraphics[scale=0.17]{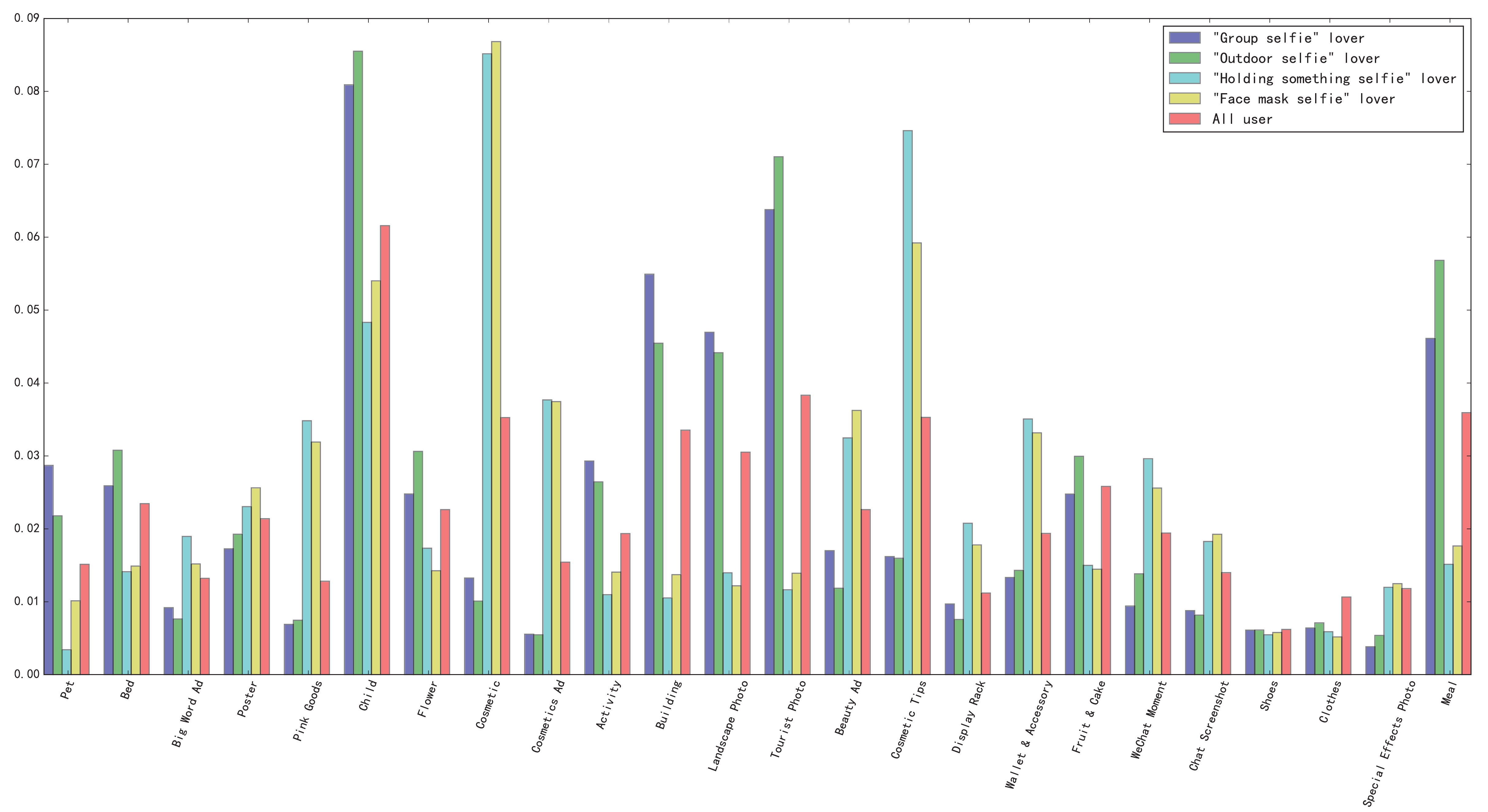}
\vspace{-8mm}
\caption{Comparison of different sets of users with special tendencies.} \label{fig:r5r6r7}\vspace{-1mm}
\end{figure}

Table~\ref{tab:r5r6r7} shows that users' posting frequency of other categories can accurately predict whether they have some typical tendencies. To further reveal the features of users with these special tendencies, for each tendency, we compute the mean value of the top 25\% users with highest tendency in the same way as Section~\ref{sec:R1-R3}. Also, we compute the mean value of all users as a comparison. Figure~\ref{fig:r5r6r7} shows the comparison of different sets of users with highest value of specific tendencies. We only select 24 typical categories to simplify the result and make the figure readable.

From Figure~\ref{fig:r5r6r7}, we can see that ``group selfie'' and ``outdoor selfie'' lovers prefer to share images that belong to Pet, Bed, Child, Activity, Large Group Photo, Building, Landscape Photo, Tourists Photo, Fruit \& Cake and Meal. On the other hand, ``face mask selfie'' lovers and ``holding something selfie'' lovers are more likely to share images that belong to Cosmetic, Cosmetics Ad, Beauty Ad, Cosmetic Tips, Chat screenshot, Special Effects Photos, Pink Goods, Display Rack and Poster. In addition, they are all uninterested in sharing images of Clothes. Also, small difference exists between ``group selfie'' and ``outdoor selfie'' lovers, for example, ``outdoor selfie'' lovers like to upload food images.

\subsubsection{Tasks R7}

\begin{figure*}[!htb]
\centering
\includegraphics[scale=0.35]{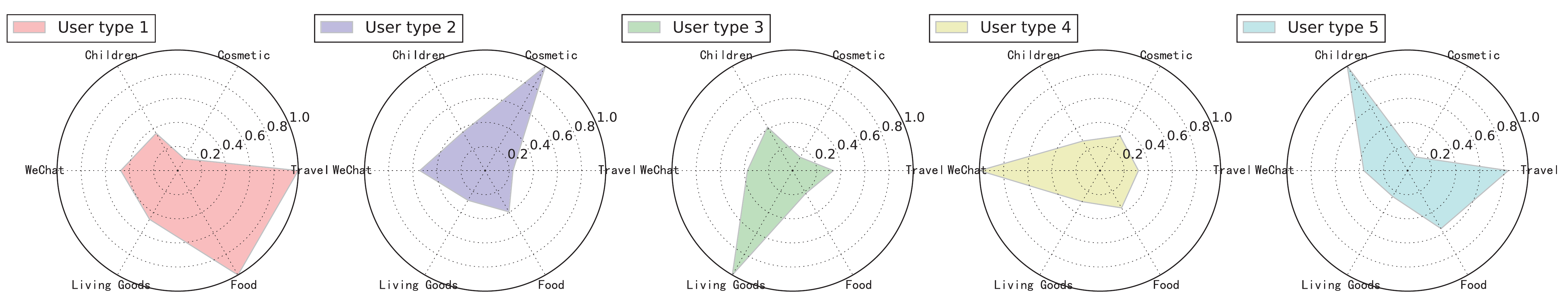}
\vspace{-5mm}
\caption{Radar plots of attributes of five user types clustered by NMF.} \label{fig:r9}\vspace{-3mm}
\end{figure*}

For this task, we first aim to cluster the users as the basis of the subsequent experiments and also as a guide to extract user's high-level attributes. To achieve user clustering, we apply Non-negative Matrix Factorization (NMF) on user's 46-dimentional frequency features. After factorization, the two non-negative matrices W and H, respectively, represent the category-type distribution and user-type distribution. We cluster users into five types and for each user we extract six high-level attributes based on the clustering result. The six attributes are Travel, Cosmetic, Children, Living Goods, WeChat and Food. Table~\ref{tab:attribute} shows the categories each attribute includes. For a user, the value of an attribute is defined as the sum of the frequencies of all categories this attribute includes. For a user type, the value of an attribute is defined as the mean attribute value of all users this user type includes. Finally, we normalize each attribute into the range  between 0 and 1 on the five user types. Figure~\ref{fig:r9} shows the radar plots of the attributes on the five user types.

\begin{table}
  \centering\small
  \caption{Six attributes and the categories included.}
  \scalebox{0.92}{
    \begin{tabular}{|c|c|}

    \hline
    Attribute & Categories Included  \\ \hline
    Travel & Landscape Photo, Tourist Photo, Building \\ \hline
    Cosmetic & Cosmetic, Cosmetics Ad, Cosmetic Tips \\ \hline
    Children & Child, Baby \\ \hline
    Living Goods & \tabincell{c}{Shoes, Clothes, Sungalss \& Handbag,\\ Necklace \& Bracelet} \\ \hline
    WeChat &\tabincell{c}{WeChat Moment, Motto, \\WeChat Expression, QR-code, WeChat Wallet, \\Chat Screenshot, Other Ad, Comic, Esaay} \\ \hline
    Food & Snack, Fruit \& Cake, Meal \\ \hline

    \end{tabular}}%
  \label{tab:attribute}%
\end{table}%

\begin{figure}[!htb]
\centering
\includegraphics[scale=0.17]{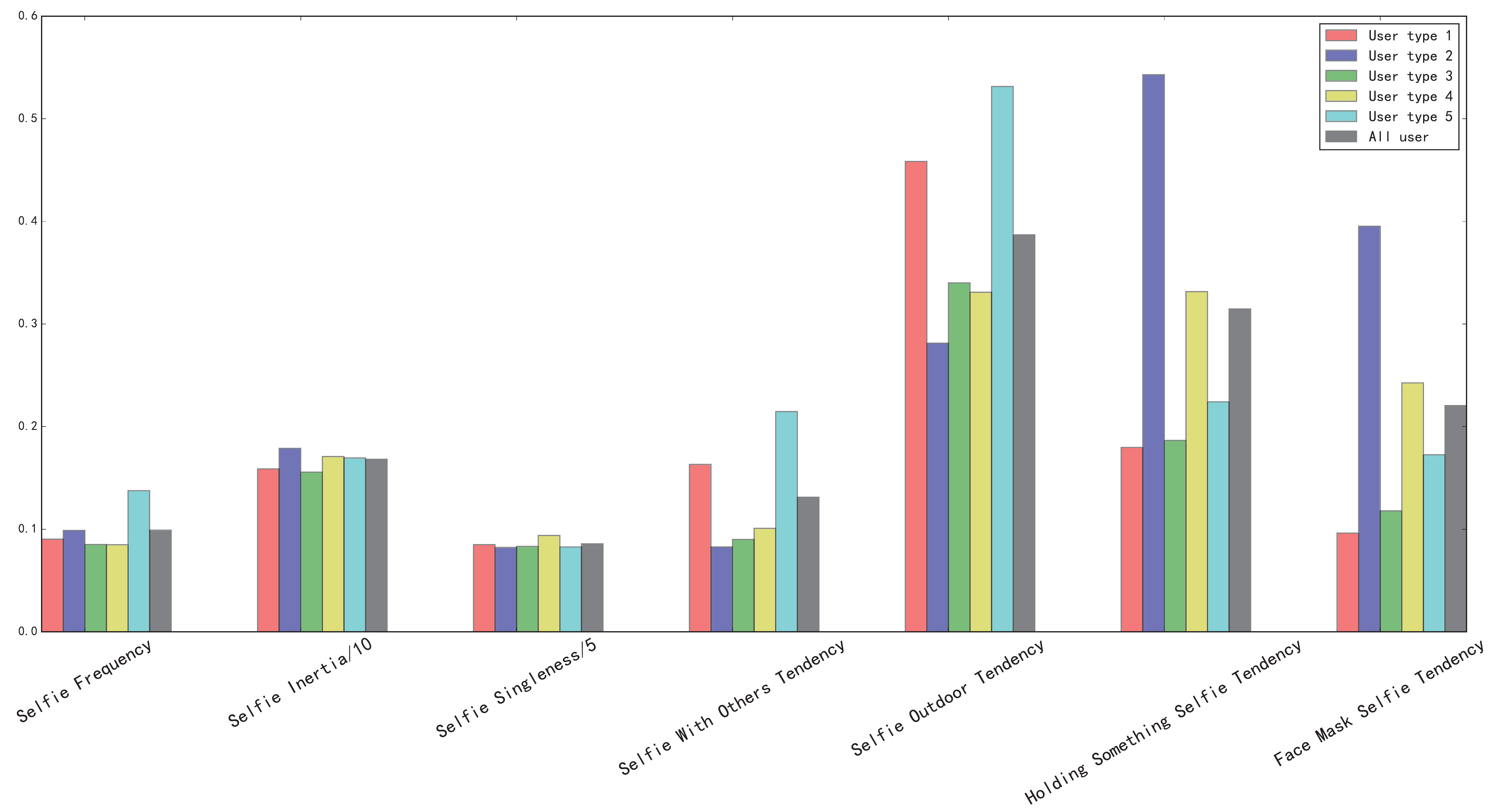}
\vspace{-7mm}
\caption{Mean values of seven selfie-posting measures among five user types.} \label{fig:r10}\vspace{-1mm}
\end{figure}

For each user type, we compute the mean value of selfie frequency, selfie inertia, selfie singleness and four advanced selfie tendencies defined in Section~\ref{sec:R5-R7} on all users this user type contains. The results are shown in Figure~\ref{fig:r10}.

It can be seen from Figure~\ref{fig:r9} that the five user types respectively reflect five types of users with different preferences. The first type of user have a preference to Travel and Food, the other four types of users respectively have a preference to Cosmetic, Living Goods, WeChat and Children (because there is high correlation between Children and Travel, so the fifth type of users also has a relatively high value of ``Travel''). We can thus determine the attribute rank based on the correlation between attribute and each selfie-posting behaviour. From Figure~\ref{fig:r10}, for selfie frequency, the correlation ranking of attributes, from high to low (also from positive to negative) is Children, Cosmetic, Travel, Food, Living Goods and WeChat. For selfie inertia, the ranking is Cosmetic, WeChat, Children, Travel, Food and Living Goods. For selfie singleness, WeChat is highest, the others are similar. For group selfie tendency, the ranking is Children, Travel, Food, WeChat, Living Goods and Cosmetic. For outdoor selfie tendency, it is Children, Travel, Food, Living Good, WeChat and Cosmetic. In the end, for both holding something selfie tendency and face mask selfie tendency, the ranking is Cosmetic, WeChat, Children, Living Goods, Travel and Food.

The ranking of the high-level attributes establishes the relation between users' selfie-posting behaviours and personal patterns in a more concise and direct way. For a set of users with a specific selfie-posting behaviour, we can easily estimate whether they have a preference for an attribute.  The results are quite reasonable. First, it is consistent with our common sense. For example, from Figure~\ref{fig:r10}, we can see that users with a preference for WeChat (user type 4) have the highest average selfie singleness, they usually post selfie without other images, which is consistent with our general knowledge that ``WeChat'' images, as their preference, is seldom posted with selfies compared with other types of images. Furthermore, this result is consistent with the reality. For instance, for users who love Travel, Food and Children (user types 1 and 5), they have high outdoor selfie tendency and group selfie tendency. This is consistent with the fact that most travel photos are taken outdoor and people love taking photos with ones who travel with them, such as their children and friends. Finally, the result is consistent with the purposes of some posted selfies. For example, users who have a preference for cosmetic (user type 2) possess the highest average holding something selfie tendency and face mask selfie tendency, this is consistent with some cosmetic lovers' intention to advertise their cosmetic products by sharing a selfie with products in their hands.

In the end, based on above conclusion that users with preferences to different high-level attributes have different selfie-posting behaviour, we further prove that user's selfie-posting behaviour can respectively predict these preferences. For each high-level attribute, we label the top 25\% users with highest value of this attribute as positive and bottom 25\% users with lowest value of this attribute as negative. Predicting each attribute can thus be regarded as a binary classification task. We use the seven selfie-posting measures as features and for each attribute we train an SVM. The 10-fold cross-validation accuracy for each task is shown in Table~\ref{tab:attrpre}. The result proves that user's selfie-posting behaviour have the potential to indicate whether a user has preference toward some attributes, which could be applied to predict significant user patterns for future research and application.

\begin{table}[htbp]
  \centering\small
  \caption{\label{tab:attrpre}10-fold cross-validation results of high-level attributes.}
  \begin{threeparttable}
    \begin{tabular}{|r|r|r|r|}

    \cline{1-2}
    High-level Attributes & Classification Result \\ \hline
    Travel  & 87.60 \\ \hline
    Cosmetic  & 89.26 \\ \hline
    Children & 76.03 \\ \hline
    Living Goods & 72.73 \\ \hline
    WeChat & 73.55 \\ \hline
    Food & 63.64 \\ \hline

    \end{tabular}%

  \end{threeparttable}
\end{table}%

\section{Conclusions and Future Work}

In this paper, we investigate deep personal patterns behind people's diverse selfie-posting behaviours. To reduce the influence of irrelevant variables and strengthen the credibility of the results, we collect a specific group of users' posted images from WeChat Moment. Based on Deep Residual Network-derived features and a clustering algorithm, we reliably classify images into different categories in an unsupervised fashion. We then characterize users by extracting three types of user-level features that reflect their interests, activities and posting habits. Furthermore, we define seven measures to comprehensively model users' selfie-posting behaviours. We predict users' selfie-posting behaviours based on three types of features. The result confirms that people's interests, activities and posting habits can help determine their selfie-posting behaviours. Moreover, comparison of typical users with specific selfie-posting behaviours clearly explains the significant personal patterns behind special selfie-posting behaviours. High-level attributes of users are established after we cluster the users by NMF. We compute the correlation ranking of attributes for each selfie-posting behaviour to measure the relation between users' selfie-posting behaviours and personal patterns in a concise and direct fashion. Finally, we design classification tasks which demonstrate the significance of user's selfie-posting behaviour on predicting user's preference toward high-level attributes. In the future, we intend to build comprehensive user profiles by fusing multimodal sensor information, including images, text, and emoji usages.

\section{acknowledgements}
We thank the support of New York State through the Goergen Institute for Data Science, and the dataset provider.

%
\bibliographystyle{abbrv}
\bibliography{sigproc}  
%
%

\end{document}